\documentclass[pdflatex,sn-mathphys-num]{sn-jnl}

\usepackage{graphicx}
\usepackage{multirow}
\usepackage{amsmath,amssymb,amsfonts}
\usepackage{amsthm}
\usepackage{mathrsfs}
\usepackage[title]{appendix}
\usepackage{xcolor}
\usepackage{textcomp}
\usepackage{manyfoot}
\usepackage{booktabs}
\usepackage{algorithm}
\usepackage{algorithmicx}
\usepackage{algpseudocode}
\usepackage{listings}
\usepackage[version=3]{mhchem}
\usepackage{upgreek}
\usepackage{gensymb}
\usepackage{array}

\begin{document}

\title[]{Geometry and electronic properties of chiral phosphorene nanotubes}

\author*[1]{\fnm{Felipe} \sur{Cervantes-Sodi}}\email{felipe.cervantes@ibero.mx}
\author[1]{\fnm{Carlos} \sur{Delmote}}
\author[2]{\fnm{José Ángel} \sur{Reyes-Retana}}

\affil[1]{\orgdiv{Department of Physics and Mathematics}, \orgname{Universidad Iberoamericana}, \orgaddress{Prolongación Paseo de la Reforma 880, Lomas de Santa Fe, 01219, \city{Ciudad de México}, \country{Mexico}}}
\affil[2]{\orgdiv{Department of Mechanics and Advanced Materials}, \orgname{Tecnológico de Monterrey}, \orgaddress{\city{Campus Santa Fe}, \country{Mexico}}}

\abstract{Phosphorene nanotubes (PNTs) exhibit highly tunable electronic properties, yet research has largely focused on high-symmetry armchair and zigzag configurations. This work presents the first systematic study of the geometric and electronic properties of chiral PNTs. Using the Chiraltube software, a comprehensive set of 214 PNTs was generated and analyzed via a 2D zone-folding mapping approach. Subsequently, a representative subset of 48 chiral PNTs was selected for \textit{ab initio} calculations covering the full chiral space. While the mapping method overestimates band gaps by directly inheriting monolayer features, it serves as a qualitative tool for specific chiral geometries—particularly armchair configurations and larger-radius tubes—by effectively capturing their subband density and dispersion profiles. Finally, this work details the evolution of structural stability and electronic properties as a function of radius and chirality, identifying clear direct-to-indirect band gap transitions driven by curvature effects, alongside metallic behavior in specific zigzag configurations.}


\maketitle

\section{Introduction}\label{sec1}
Since the early days of nanotechnology, nanotubes (NT) have become the archetypal nanomaterial, represented primarily by carbon nanotubes~\cite{dresselhaus1998physical}. However, with the growing emergence of a plethora of nanotubes composed of other elements and compounds, such as MoS$_2$~\cite{huang2025atlas}, WS$_2$~\cite{tenne2010recent}, phosphorene~\cite{guo2014phosphorene}, boron nitride~\cite{chopra1995boron}, and silicene~\cite{kara2012review}, among many others, establishing a general approach to the generation of their atomic coordinates and their subsequent study has become a priority. 

Typically, within a computational perspective, these NTs are generated by rolling up a two-dimensional monolayer (ML) along a specific rolling direction, uniquely defined by a pair of chiral indices $(n, m)$ \cite{dresselhaus1998physical,de2024chiraltube}. 

Once generated, each $\text{NT}(n, m)$ can be studied using various computational approaches to determine its mechanical, electronic, and optical properties

Phosphorene nanotubes (PNTs) can be derived from any of the phosphorus allotropes, \textit{i.e.}, $\alpha$-P, $\beta$-P, $\gamma$-P, and $\delta$-P. Among these, $\alpha$-P, also known as black phosphorus (BP), exhibits an orthorhombic lattice and remains the most studied due to its thermodynamic stability\cite{zhang2024recent, bonometti2025toward}. While bulk BP possesses a narrow band gap of $\sim 0.31$~eV~\cite{maruyama1981synthesis}, monolayer black phosphorene (MLBP) displays a significantly larger direct band gap, experimentally measured at $\sim 2$~eV~\cite{liu2014phosphorene, li2014black}. Theoretical calculations of MLBP yield values ranging from $\sim 0.8$ to $1.0$~eV within standard density functional theory (DFT), and up to $2.3$~eV when incorporating $GW$ many-body corrections~\cite{allec2016inconsistencies, ccakir2015significant, zhang2024recent, liu2014phosphorene}. This significant widening of the band gap from the bulk to the ML limit is mainly attributed to the quantum confinement effect, which fundamentally modify the electronic band structure upon its dimensional reduction~\cite{chaudhary2022phosphorene}.

Recently, phosphorene-based nanostructures, such as nanoribbons, nanotubes, and van der Waals multilayers, have seen significant progress in their synthesis and characterization, opening new avenues for technological applications~\cite{guo2014phosphorene, akhtar2017recent, sorkin2017recent}. Indeed, these materials have already been integrated into high-performance transistors, electrodes, and solar cells~\cite{keyes1953electrical, Abernethy2003}. Regarding phosphorene nanotubes, several studies have been conducted, focusing primarily on highly symmetric armchair (0, $m$) and zigzag ($n$, 0) configurations. Some research has also explored other high-symmetry chiral tubes, as these are computationally accessible and involve a relatively low number of atoms. DFT calculations reveal that PNTs undergo a direct-to-indirect band gap transition as the nanotube's diameter decreases, a phenomenon largely driven by structural strain effects~\cite{allec2016inconsistencies}. Typically, armchair PNTs exhibit semiconducting properties with narrow direct band gaps~\cite{guo2014phosphorene,allec2016inconsistencies}, making them promising candidates for applications in nanoelectronics and optoelectronics. This behavior stands in contrast to the metallic conductivity observed in zigzag PNTs for low $n$ index ~\cite{fernandez2016chiral}.

Although some specific electronic properties are well-documented for high-symmetry zigzag and armchair PNTs, a more generalized understanding of their electronic behavior can be achieved through band mapping. A rough estimate of a nanotube's band gap can be obtained by mapping the 2D Brillouin zone of phosphorene onto the 1D reciprocal space of an unfolded chiral nanotube. This zone-folding procedure was originally formalized for carbon nanotubes~\cite{samsonidze2003concept}, demonstrating that the metallic or semiconducting character of a CNT depends on the relative position of its ``cutting lines'' with respect to the $K$ and $K'$ points in the 2D reciprocal space. Recently, this methodology was applied to create an atlas of MoS$_2$ chiral nanotubes~\cite{huang2025atlas}, providing initial estimates of electronic properties that were later refined using deep learning-enhanced DFT calculations.

The primary contribution of this work is the first systematic study of truly chiral phosphorene nanotubes. To achieve this, we first utilized the Chiraltube package~\cite{de2024chiraltube} to generate a comprehensive set of 214 nanotubes, determining their geometric structures and reciprocal lattice vectors. Following this structural characterization, we applied the zone-folding procedure to map the cutting lines of the unfolded nanotubes onto the 2D Brillouin zone of phosphorene, providing a predictive framework for their 1D electronic properties.

Subsequently, a representative subset of 48 PNTs encompassing a diverse range of chiralities was investigated via \textit{ab initio} calculations. This study involved full geometry optimizations and band structure calculations. This selection of nanotubes was designed to cover a broad structural spectrum, including zigzag ($n$, 0), armchair (0, $n$), and ladder ($n, n$)—to general chiral nanotubes, including (2, 8), (5, 6), and (6, 12).

\begin{figure}[t]
\centering
\includegraphics[width=0.6\textwidth]{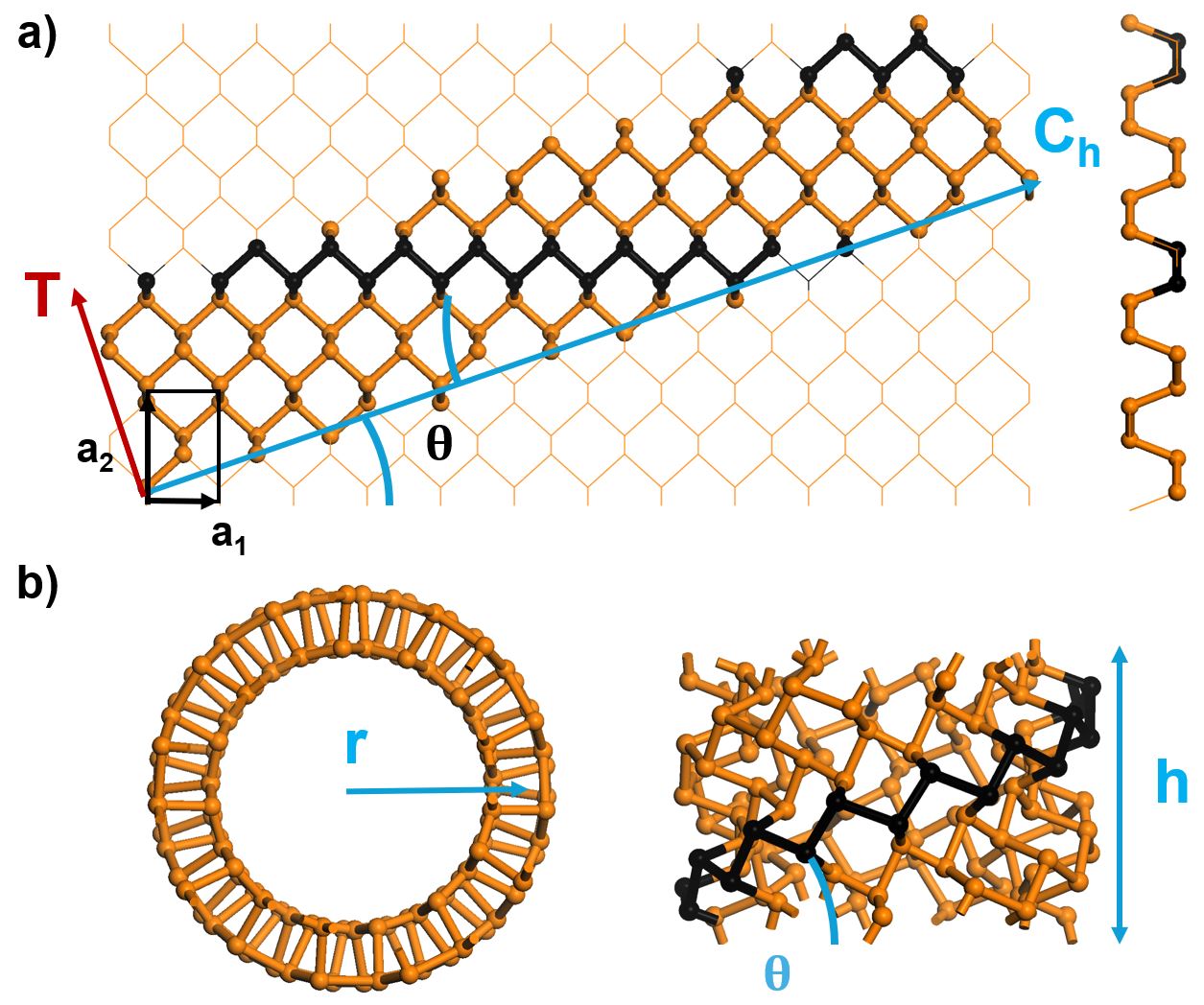}
\caption{\label{Structures} Ball-and-stick representation of the black phosphorene monolayer and the resulting nanotube. (a) 2D phosphorene in the $xy$ and $zy$ projections, where $\mathbf{a}_1$ and $\mathbf{a}_2$ represent the lattice vectors of the unit cell. The chiral vector ($\mathbf{C_h}$), translational vector ($\mathbf{T}$), and chiral angle ($\theta$) for the specific (12, 3) PNT are indicated. The row of atoms highlighted in black serves as a visualization guide for the rolling process. (b) Top and side views of the relaxed (12, 3) PNT showing the tubre radius $r$, the nanotube height $h$, $\theta$.}
\end{figure}

\section{Results and discussion}\label{sec2}

Figure \ref{Structures}a presents the characteristic puckered structure of the black phosphorene monolayer\cite{zhang2024recent}. To illustrate the construction process, the formation of a PNT with chiral indices (12, 3) is shown. The chiral vector ($\mathbf{C_h}$), the translational vector ($\mathbf{T}$), and the chiral angle ($\theta$) are indicated. In the 2D lattice, we highlight the specific ribbon of atoms that, upon rolling, constitutes the nanotube. The atoms highlighted in black provide a clear visualization of how the 2D orientation transforms into the 1D chiral morphology, preserving the relative orientation between the phosphorene axes and the tube's longitudinal axis.

The geometric properties of these nanotubes (Supplementary Table~1S) are calculated directly by Chiraltube using generalized formulas adapted from CNT theory~\cite{samsonidze2003concept, de2024chiraltube}, as summarized in Table~\ref{Table}. 

\begin{table}[h!]
\centering
\caption{Parameters for phosphorene nanotubes with chiral index $(n, m)$.}
\begin{tabular}{llll}
\toprule
Symbol & Parameters & Formula & Value/Property \\ \midrule
$\mathbf{a}_1, \mathbf{a}_2$ & Lattice constants & $\begin{aligned} \mathbf{a}_1 &= 3.30~\text{\AA} \\ \mathbf{a}_2 &= 4.62~\text{\AA} \end{aligned}$ & From DFT \\[6pt] \hline
$n$, $m$ & Chiral indices & -- & Integers \\ 
& & & $0 \le m \le n$ \\ \hline
$\mathbf{C}_h$ & Chiral vector & $\mathbf{C}_h = n\mathbf{a}_1 + m\mathbf{a}_2 \equiv (n, m)$ & $x, y$ coordinates \\ 
$L$ & Length of $\mathbf{C}_h$ & $L = \sqrt{(n a_1)^2 + (m a_2)^2}$ & Real \\ \hline
$r_0$ & Unrelaxed tube radius & $r_0 = \dfrac{L}{2\pi}$ & Real \\ 
$r$ & Relaxed tube radius & $r$ & From DFT \\ \hline
$\theta$ & Chiral angle & $\tan \theta = \dfrac{m a_2}{n a_1}$ & $0 \le \theta \le \dfrac{\pi}{2}$ \\ \hline
$t_1, t_2$ & Parameters for translational vector & -- & $\dfrac{t_1}{t_2} = \dfrac{n\left|\mathbf{a}_1\right|}{m\left|\mathbf{a}_2\right|}$ \\[6pt] \hline
$\mathbf{T}$ & Translational vector & $\mathbf{T} = t_1\mathbf{a}_1 + t_2\mathbf{a}_2 \equiv (t_1, t_2)$ & $x, y$ coordinates \\ 
$T$ & Length of $\mathbf{T}$ & $T = \sqrt{(t_1 a_1)^2 + (t_2 a_2)^2}$ & Real \\ \hline
$h_0$ & Unrelaxed tube height & $h_0 = T$ & Real \\ 
$h$ & Relaxed tube height & $h$ & From DFT \\ \hline
$N_C$ & 2D unit cell per NT supercell & $N_C = \dfrac{L \cdot T}{a_1 \cdot a_2}$ & Integer \\ \hline
$N$ & Number of atoms per supercell & $N = 4 N_C$ & Integer \\ \midrule
$\mathbf{b}_1, \mathbf{b}_2$ & Reciprocal lattice vectors & $\mathbf{b}_1 = \left(\dfrac{2\pi}{a_1}, 0\right), \mathbf{b}_2 = \left(0, \dfrac{2\pi}{a_2}\right)$ & $x, y$ coordinates \\ \hline
$\mathbf{K}_1, \mathbf{K}_2$ & Reciprocal vectors & $\begin{aligned} \mathbf{K}_1 &= \dfrac{1}{N}(-t_2\mathbf{b}_1 + t_1\mathbf{b}_2) \\ \mathbf{K}_2 &= \dfrac{1}{N}(m\mathbf{b}_1 - n\mathbf{b}_2) \end{aligned}$ & $x, y$ coordinates \\[10pt] \hline
$N_{\mathrm{CL}}$ & Number of cutting lines & $N_{\mathrm{CL}} = N_C$ & Integer \\
\bottomrule
\end{tabular}
\label{Table}
\end{table}

\begin{figure}[h]
\centering
\includegraphics[width=1\textwidth]{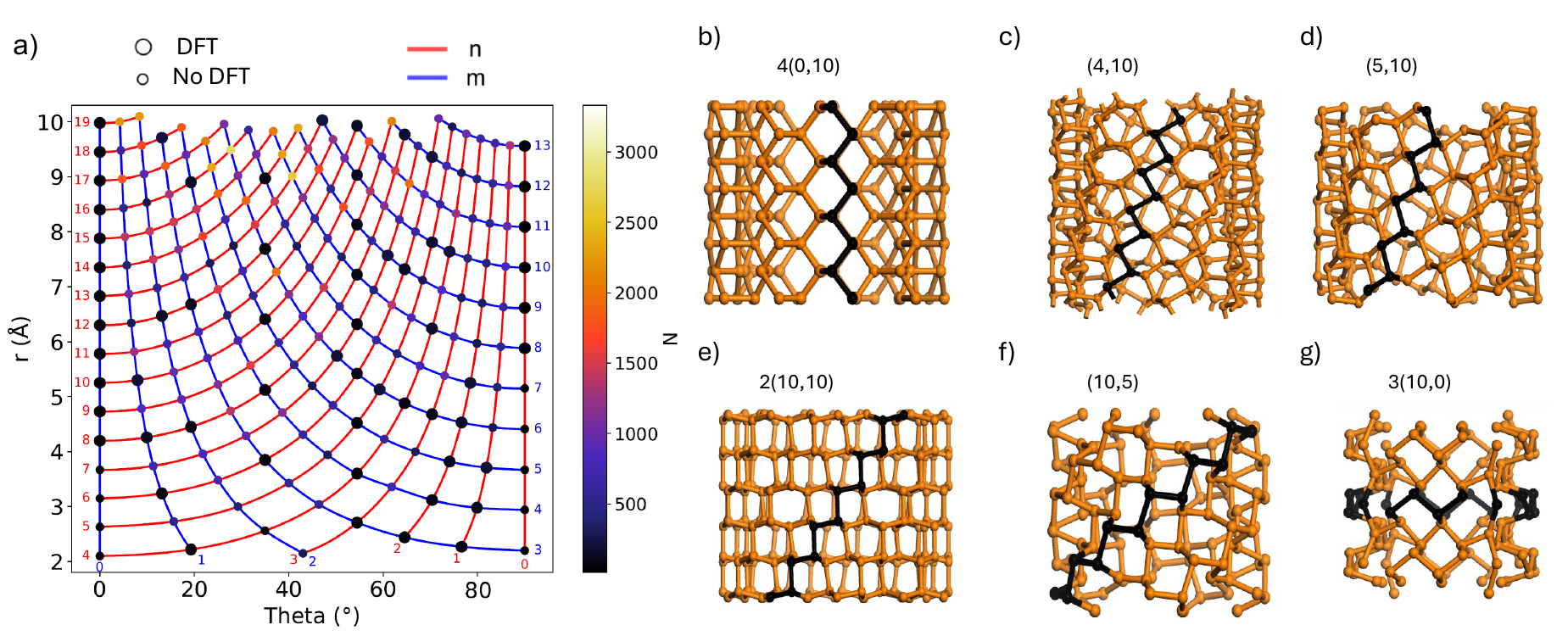}
\caption{\label{R_theta_6tubes} 
Structural classification of PNTs. (a) Radius ($r$) vs. chiral angle ($\theta$) vs. number of atoms ($N$) for the 214 generated nanotubes. Each point represents an ($n$, $m$) configuration, with red and blue lines tracing constant $n$ and $m$ series, respectively. The color scale indicates the number of atoms $N$ in the PNT supercell, with the large black dots highlighting the 48 structures selected for \textit{ab initio}  study (DFT), while smaller markers indicate the remaining generated geometries (No DFT). (b--g) Atomic representations of representative nanotubes with a shared index of 10, including (b) armchair (0, 10), (c) chiral (4, 10), (d) chiral (5, 10), (e) ladder (10, 10), (f) chiral (10, 5), and (g) zigzag (10, 0). Black atoms highlight the transformation of an equivalent zigzag row across different chiral angles. For clarity, panels (b), (e), and (g) display 4, 2, and 3 supercells, respectively.}
\end{figure}

In this way, a total of 214 PNTs were generated by Chiraltube~\cite{de2024chiraltube}, and classified as armchair (0, $m$), zigzag ($n$, $0$), ladder ($n$, $n$) or chiral ($n$, $m$). Figure~\ref{R_theta_6tubes}a presents the distribution of the initial radius $r_0$ as a function of the chiral angle $\theta$. The blue and red curves illustrate the continuous evolution of the chiral indices along constant $m$ and $n$ series, respectively, where the chiral angle ranges from 0$^\circ$ for $(n, 0)$ zigzag nanotubes to 90$^\circ$ for $(0, m)$ armchair tubes. 

Each chirality is represented by a color coded circle whose color reflects the number of atoms ($N$) in the supercell, following a scale from black and purple to blue, red, orange, and yellow. This map reveals a drastic increase in $N$ for general chiral tubes—particularly in intermediate chiral angle regions—where $N$ can exceed several thousands. This color code highlights the severe computational challenge of performing \textit{ab initio} calculations on truly chiral PNTs. Specifically, the subset of 48 PNTs selected for DFT calculations comprises structures with a low atom count ($N < 236$) located at the black end of the scale, and they are highlighted with larger circle radii. Conversely, smaller circles with purple-to-yellow colors represent the remaining PNTs with a high number of atoms, which were excluded from the \textit{ab initio} analysis.


Among the various families, the $(0, m)$ series exhibits the smallest unit cells, followed by the $(n, 0)$, $(2n, n)$, $(n, m)$, $(4n, n)$, $(2n, 3n)$, and the remaining chiral nanotube configurations. 

Figures~\ref{R_theta_6tubes}b--g provide a visual comparison of representative nanotubes sharing a common index of 10. By tracking a specific zigzag row of atoms (highlighted in black), the structural anisotropy of phosphorene is clearly illustrated across different chiralities, showing how the orientation shifts relative to the tube axis as $\theta$ varies.

To initiate the first-principles investigation, geometry optimizations, formation energy evaluations, and electronic structure calculations of the MLBP baseline were performed to benchmark our model against existing literature and provide an optimized unit cell for constructing the nanotubes. The relaxed lattice parameters for bulk-equivalent phosphorene were found to be $a_1 = 3.299$~\AA\ and $a_2 = 4.619$~\AA\ (Figure~\ref{Structures}a). The calculated P--P interatomic bond lengths are $2.22$~\AA\ and $2.26$~\AA, whereas the bond angles defined by three consecutive atoms are $\alpha = 95.9^\circ$ and $\beta = 104.2^\circ$. These structural results show excellent agreement with previously reported data~\cite{fernandez2016chiral, fernandez2019structural, qiao2014high, zhang2024recent}.

\begin{figure}[h]
\centering
\includegraphics[width=0.9\textwidth]{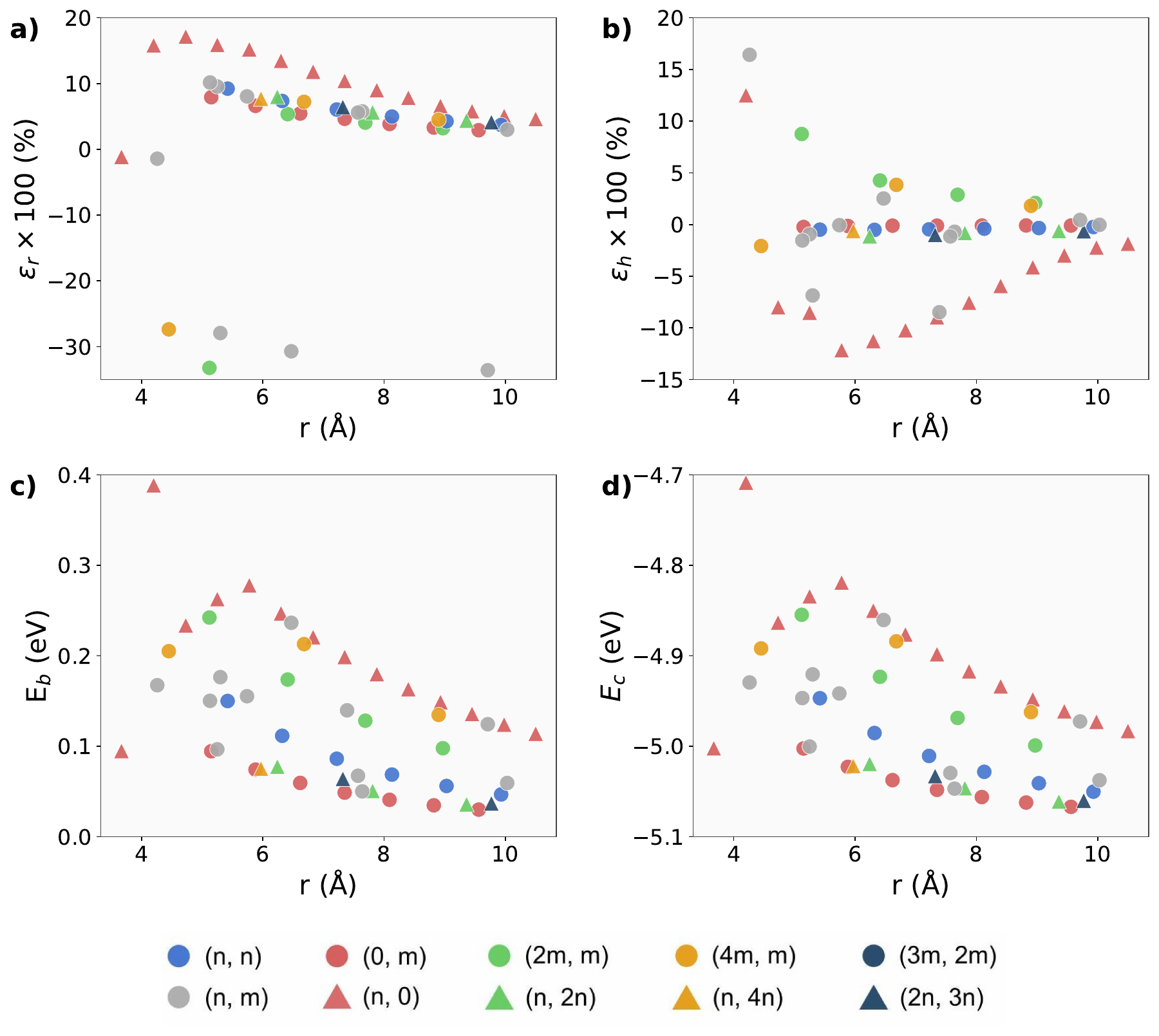}
\caption{\label{geometry} 
Structural relaxation and energetic stability of PNTs as a function of the relaxed tube radius $r$. (a) Percentage radial strain ($\varepsilon_r \times 100$), (b) percentage longitudinal strain ($\varepsilon_h \times 100$), (c) strain/binding energy per atom ($E_b$), and (d) cohesive energy per atom ($E_c$) of the relaxed nanotubes. Color and symbol codes identify the different chiral families as indicated in the legend}
\end{figure}

Due to the puckered double-layer structure of phosphorene, the nanotubes generated via Chiraltube inherently store significant radial strain ($\varepsilon_r$), characterized by the compression of the innermost atomic layer and the tension of the outermost one~\cite{li2018tunable, allec2016inconsistencies}. Additionally, a longitudinal (axial) strain ($\varepsilon_h$) emerges along the nanotube axis\cite{fernandez2019structural, sorkin2017recent}. Consequently, a full geometry optimization is essential to relax these internal stresses and attain the local energy minimum. After relaxation, the radial and longitudinal strains are evaluated as $\varepsilon_r = (r/r_0) - 1$ and $\varepsilon_h = (h/h_0) - 1$, respectively.

Figures~\ref{geometry}a and~\ref{geometry}b show the evolution of these strains as a function of the relaxed radius. The $(n, 0)$ series presents the largest radial strain values alongside the lowest (more negative) values for $\varepsilon_h$, effectively defining the upper and lower boundaries of the structural deformation spectrum. In contrast, families such as $(0, m)$ and $(n, 2n)$ define the lower limit for radial strain. Notably, several families exhibit almost zero longitudinal strain ($\varepsilon_h \approx 0$), specifically the $(n, n)$, $(0, m)$, $(n, 2n)$, and $(n, 4n)$ series. On the other hand, nanotubes in the $(2m, m)$ and $(4m, m)$ series represent the upper boundary for $\varepsilon_h$. General chiral tubes (plotted in gray) behave as expected, with their strain values systematically oscillating between the limits set by the armchair and zigzag configurations.

Nanotubes with small radii are characterized by high initial stress and intense interatomic forces, which often trigger structural transitions toward non-circular cross-sections. For instance, we observed the formation of oval $(7, 0)$, rectangular $(9, 0)$ and square $(10, 0)$ cross-sections (Supplementary Fig.~S1). Also, when the optimization for these zigzag tubes starts from a slightly asymmetric configuration, they tend to relax into more stable, equivalent armchair-like structures. Other non-zigzag tubes also exhibit non-circular geometries, such as the irregular $(8, 1)$, rectangular $(8, 2)$ and triangular $(10, 1)$ configurations.

PNTs $(8, 2)$, $(8, 4)$, $(10 1)$, $(12, 2)$, and $(18, 3)$ are non-zigzag and non-armchair nanotubes that exhibit large negative values for $\varepsilon_h$ (Figure~\ref{geometry}b). After full geometry relaxation, these real chiral nanotubes possess final diameters significantly smaller than those of the as-generated tubes. For instance, the $(8, 2)$ PNT starts with an extremely narrow initial radius of $4.45$~\AA{} (within the stability limit), but upon geometry optimization, it loses its circular cross-section and adopts a rectangular one with an average radius of $3.23$~\AA. Similarly, the $(8, 4)$, $(12, 2)$, and $(18, 3)$ PNTs start with radii of $5.12$, $6.47$, and $9.71$~\AA{}, respectively, and shrink after optimization to $3.41$, $4.48$, and $6.45$~\AA{}. These small-diameter nanotubes preserve their circular cross-sections and initial chiral symmetry following optimization.

Chiraltube operates within specific error thresholds when searching for translational indices and folding the 2D material into a nanotube. All nanotubes in this work were generated using the default error tolerance of $0.05\%$. While adopting a stricter error tolerance could reduce the discrepancy between the initial and relaxed radii, this approach is impractical because smaller error values lead to excessively large unit cells. For example, the $(18, 3)$ PNT comprises $228$ atoms under the default error, but this count surges to $3,720$ atoms for a smaller error threshold, making the structure computationally unmanageable for DFT calculations.

We consider that these nanotubes serve as a rigorous test for Chiraltube, demonstrating its capability to generate nontrivial nanotubes of 2D materials featuring complex bilayer structures and a tunable number of atoms in their unit cells. Other works have highlighted the critical relevance of strain—whether residual or induced—in tailoring the electronic properties of PNTs\cite{li2018tunable, sorkin2017recent, allec2016inconsistencies}.

The stability of the relaxed nanotubes was evaluated by calculating the binding energies ($E_b$) and cohesive energies ($E_c$) per atom:

\begin{equation}
\label{eq:binding}
E_b = \frac{E_T}{N} - \frac{E_{\mathrm{ML}}}{4}
\end{equation}
and
\begin{equation}
\label{eq:cohesive}
E_c = \frac{E_T - N E_P}{N}
\end{equation}

\noindent where $E_T$ is the total energy of the nanotube with $N$ atoms, while $E_{\mathrm{ML}}$ and $E_P$ are the energies of the phosphorene monolayer unit cell and a single isolated phosphorus atom, respectively.

MLBP exhibits a cohesive energy of $-5.09$~\text{eV}, confirming its stability as a 2D sheet in agreement with previous studies~\cite{durajski2021influence, li2018tunable}. As anticipated, all evaluated nanotubes display positive $E_b$ values (Figure~\ref{geometry}c), reflecting the higher stability of the uncurved 2D monolayer relative to the 1D tubular configurations~\cite{fernandez2019structural, li2018tunable}. Conversely, the cohesive energies $E_c$ remain strictly negative and asymptotically converge toward the monolayer benchmark as the tube radius expands (Figure~\ref{geometry}d). The thermodynamic stability across chiral families follows a well-defined progression: $\text{zigzag } (n, 0) < (4m, m) < (2m, m) < (n, n) < (2n, 3n) < (n, 2n) < (n, 4n) < \text{armchair } (0, m)$,
confirming that armchair $(0, m)$ PNTs constitute the energetically most favorable configurations~\cite{fernandez2016chiral, li2018tunable} (complete numerical datasets are compiled in Supplementary Table S1).

Furthermore, for nanotubes undergoing substantial axial contraction (large negative $\varepsilon_h$), their calculated $E_b$ and $E_c$ values span intermediate ranges. Specifically, the $(12, 2)$ and $(18, 3)$ chiral PNTs exhibit some of the highest $E_b$ values, indicating that they belong to the least thermodynamically stable configurations, closely approaching the high-energy threshold of the zigzag family.

\begin{figure}[t]
\centering
\includegraphics[width=.8\textwidth]{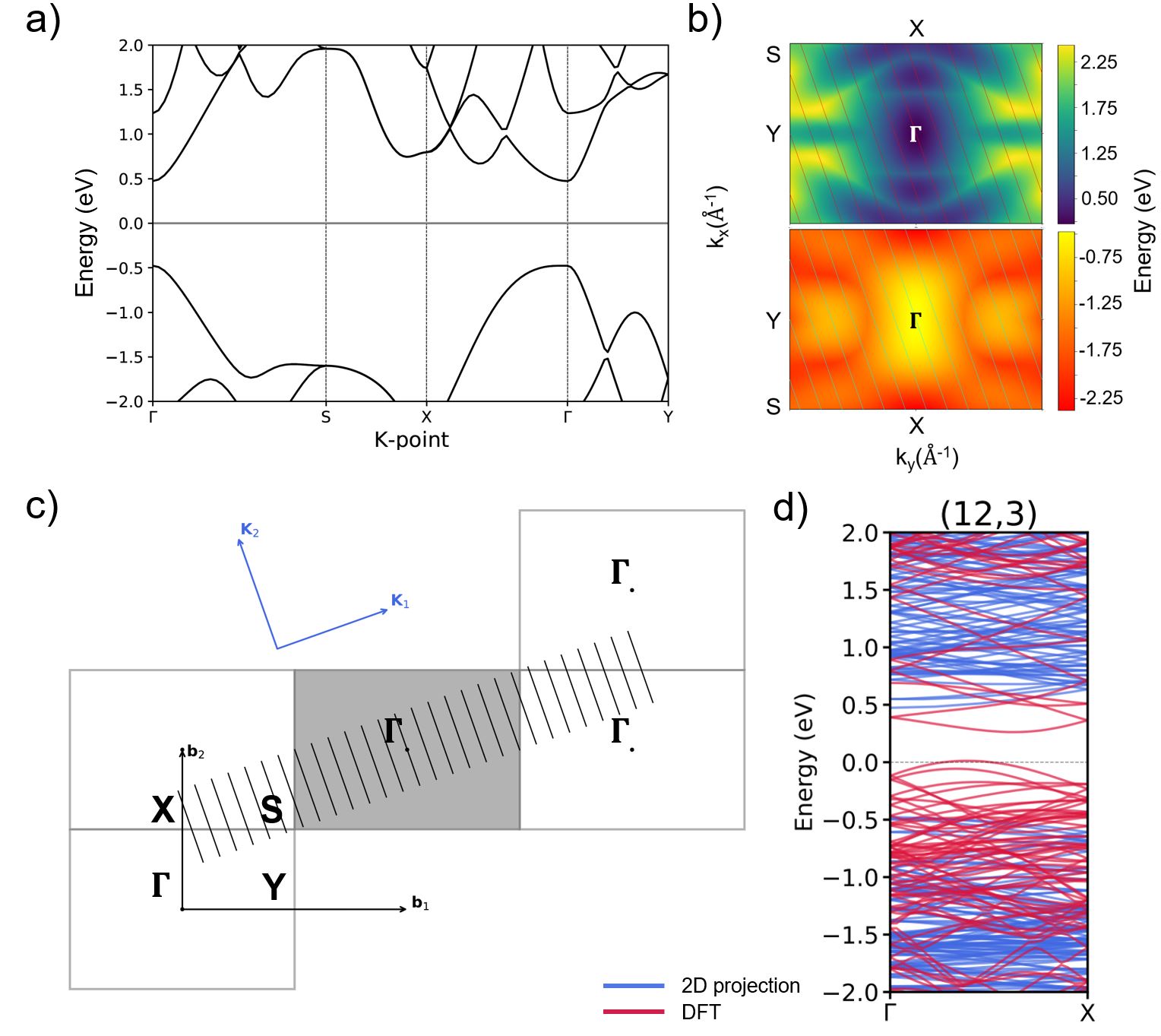}
\caption{\label{PNTS-Bands}(a) Electronic band structure of monolayer black phosphorene with high-symmetry points, alongside (b) 2D color plots of the highest valence band and lowest conduction band. (c) Reciprocal lattice space spanning five complete Brillouin zones, showing the reciprocal lattice vectors of 2D phosphorene ($\mathbf{b}_1$, $\mathbf{b}_2$) and those of the $(12,3)$ PNT ($\mathbf{K}_1$, $\mathbf{K}_2$). The 27 cutting lines are depicted in the fully extended representation in (c) and mapped in the fully reduced representation across the bands in (b). (d) Comparison of the band structure for the $(12,3)$ PNT obtained via the 2D projection model (blue lines) and full DFT calculations (red lines).}
\end{figure}

The band structure of MLBP is displayed in Fig.~\ref{PNTS-Bands}a, alongside 2D color plots of the highest valence and lowest conduction bands highlighting the Brillouin zone high-symmetry points (Fig.~\ref{PNTS-Bands}b). The direct band gap at the $\Gamma$ point is found to be $0.95$~eV at the PBE level of theory. Although it is well known that standard generalized gradient approximations (GGA-PBE) underestimate band gaps compared to hybrid functionals (\textit{e.g.}, HSE06) or many-body perturbation theory ($GW$), both PBE and HSE06 yield remarkably similar trends regarding electronic features—such as effective masses and strain-induced variations—in both MLBP and PNTs~\cite{sa2015electronic, fernandez2016chiral, cai2014layer, guo2014phosphorene, ccakir2015significant}. Therefore, utilizing the PBE functional is computationally pragmatic and fully justified for this study, where the primary objective is to establish general physical and electronic trends. Consequently, all electronic band structures for MLBP and PNTs presented herein are calculated at the PBE level. Quantitatively, the calculated effective masses for electrons are $0.16\,m_e$ and $0.65\,m_e$ along the Y and X directions, respectively, whereas for holes they are $0.14\,m_e$ and $8.06\,m_e$ along the same directions. The heavy, light, or ultralight character of these effective masses is in close agreement with values reported in the literature~\cite{peng2014strain, sa2015electronic, ccakir2015significant}.

Once the DFT calculation of MLBP is completed, mapping the cutting lines of an unfolded nanotube onto the reciprocal lattice of the 2D monolayer allows, in some cases, a first-order prediction of the overall band structure profile of the nanotubes.

Figure~\ref{PNTS-Bands}c illustrates the 2D reciprocal lattice including the reciprocal lattice vectors of MLBP ($\mathbf{b}_1, \mathbf{b}_2$) alongside an example of the reciprocal lattice vectors for an unfolded PNT ($\mathbf{K}_1, \mathbf{K}_2$), using the $(12, 3)$ PNT as a representative case. For clarity, $\mathbf{K}_1$ and $\mathbf{K}_2$ are enlarged and highlighted in blue. The 27 parallel and equidistant line segments constitute the $\mathbf{K}_2$-extended representation of the cutting lines for the $(12, 3)$ nanotube. These lines run parallel to $\mathbf{K}_2$, starting at $\Gamma$, and are periodically repeated 27 times with displacements of $\pm \mathbf{K}_1 N_{\text{CL}}/2$.

Figure~\ref{PNTS-Bands}b shows the fully reduced representation of these 27 cutting lines for the $(12, 3)$ PNT projected onto the highest valence band and lowest conduction band of MLBP. Extracting the energy values along these trajectories yields the band structure of the unfolded nanotube model, plotted in blue in Fig.~\ref{PNTS-Bands}d alongside the full DFT band structure in red.

Because MLBP possesses a direct band gap at the $\Gamma$ point, during the mapping process, by construction, we necessarily project the same band gap in the 2D projection over the unfolded 2D PNTs. As a consequence, the direct gap of the parent 2D material is always inherited by the nanotubes in this framework, yielding a gap of $0.95$~\text{eV} for all PNTs. 

Comparing this result with full DFT calculations reveals a significant overestimation of the band gap in the 2D zone-folding mapping model (\textit{e.g.}, $0.95$~\text{eV} projected versus $0.25$~\text{eV} from DFT for the $(12, 3)$ PNT). This overestimation occurs systematically across all evaluated nanotubes due to the neglect of curvature-induced orbital rehybridization and strain effects.

Nevertheless, the overall dispersion profile of the bands is remarkably well preserved for several families and individual nanotubes. Supplementary Fig.~S1 presents the complete set of band structures obtained via both band mapping and direct DFT calculations. These comparisons demonstrate that while the band gap magnitude is modified, the dispersion and overall subband density are maintained for the $(0, m)$ armchair and $(n, n)$ ladder families, as well as for specific chiral nanotubes such as $(4, 10)$, $(5, 6)$, $(7, 6)$, and $(8, 1)$.

In general, the band density resulting from the folding of the reciprocal space—especially for larger nanotubes characterized by a reduced reciprocal unit cell—is largely preserved when comparing the zone-folding mapping results with full DFT calculations.

Throughout the reciprocal space, the cutting lines of different PNTs define distinct trajectories, resulting in unique band density profiles for each chiral configuration.

\begin{figure}[t]
\centering
\includegraphics[width=1\textwidth]{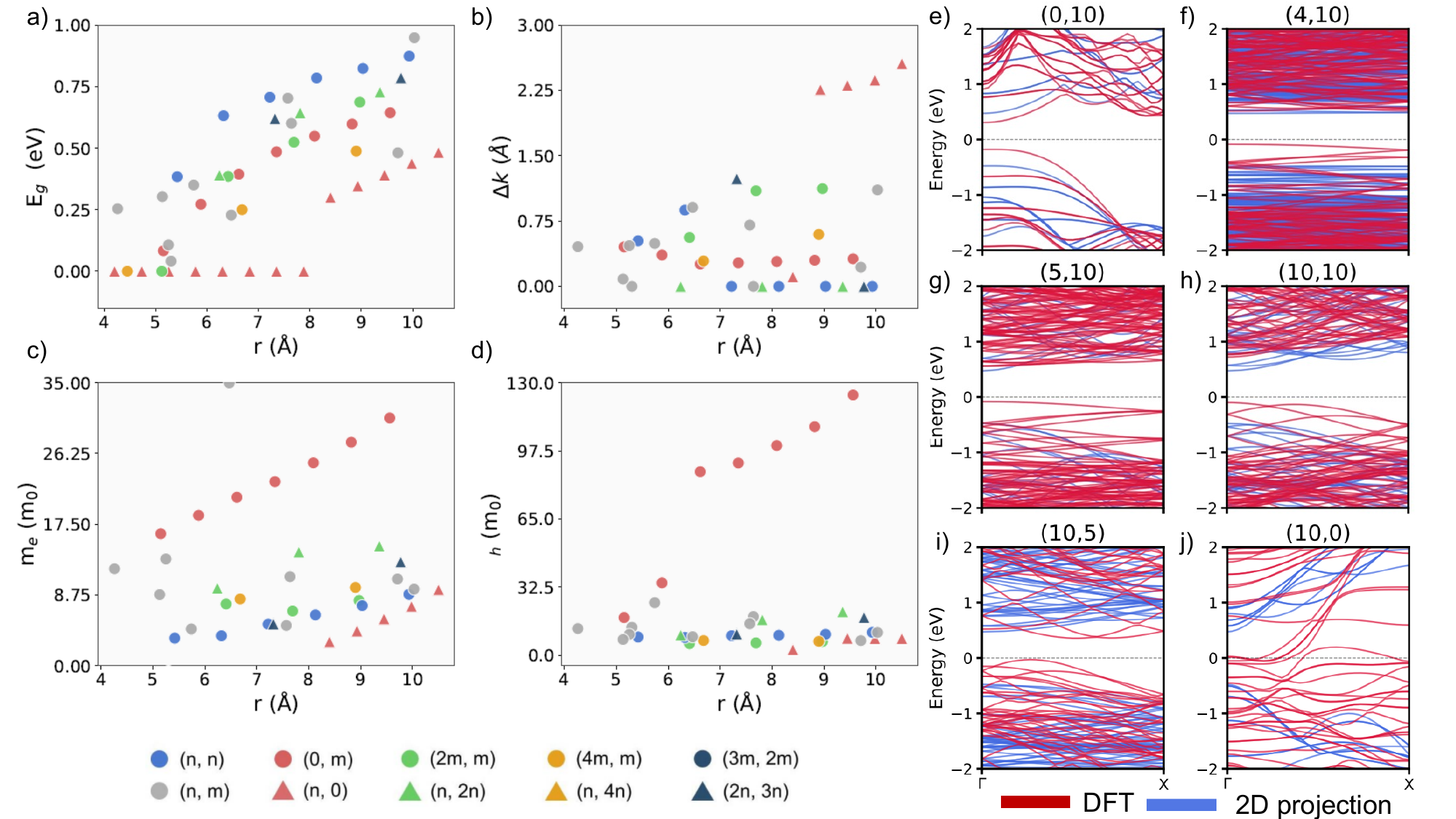}
\caption{\label{Eg-DFT} (a) Band gap as a function of radius obtained by DFT, (b) momentum transfer vector $\Delta k$, (c) electron effective mass $m_e$, and (d) hole effective mass $m_h$. Panels (e)--(j) present the band structure comparisons for representative PNTs: (e) $(0, 10)$, (f) $(4, 10)$, (g) $(5, 10)$, (h) $(10, 10)$, (i) $(10, 5)$, and (j) $(10, 0)$. Blue lines represent the mapped bands derived from the unfolded 2D monolayer, whereas red lines denote the full DFT calculated bands.}
\end{figure}

Focusing on DFT results, quantum confinement and nanotube curvature significantly modify the band structure~\cite{fernandez2016chiral, fernandez2019structural, allec2016inconsistencies, guo2014phosphorene}, establishing a clear trend in the band gap as a function of tube radius and chirality. Figures~\ref{Eg-DFT}a–d show the band gap, momentum transfer vector ($\Delta k = k_{\mathrm{VBM}} - k_{\mathrm{CBM}}$), and effective masses obtained via DFT. As expected and consistent with previous literature, the band gap decreases as the PNT radius diminishes. Furthermore, there is a strong dependence of the band gap on nanotube chirality: zigzag configurations establish the lower limit for $E_g$, whereas the $(n, n)$ lader family exhibits the highest values. Armchair nanotubes feature direct band gaps with values close to the indirect gaps of the $(2n, n)$ family, lying above the indirect gaps of the $(4n, n)$ series.

Nanotubes with radii around $5$~\AA\ and zigzag PNTs ranging from $(8, 0)$ to $(15, 0)$ exhibit metallic behavior with a zero band gap (in contrast to Ref.~\cite{fernandez2016chiral}, which reported the $(10, 0)$ and $(11, 0)$ zigzag nanotubes as semiconductors). The $(8, 1)$ nanotube stands as an exception, displaying a finite band gap of $0.25$~\text{eV}; its unrelaxed radius is $4.36$~\AA, and its relaxed structure departs from a circular cross-section, adopting an irregular geometry that likely triggers the opening of the band gap.

Interestingly, the $(n, n)$ family, featuring a chiral angle close to $45^\circ$, presents a relatively large band gap in agreement with Ref.~\cite{fernandez2016chiral}, exhibiting direct band gaps for $n = 6$ and $7$, and indirect ones for $n = 8$--$11$. The largest band gap corresponds to the $(13, 10)$ nanotube, which possesses a chiral angle of $47.12^\circ$, closest to $45^\circ$. Although the $(7, 5)$ nanotube was not calculated due to the exceptionally large number of atoms in its unit cell, we anticipate it could exhibit an even larger band gap given its chiral angle of $45.002^\circ$ (all band gap values are compiled in Supplementary Table S1).

Figure~\ref{Eg-DFT}b shows a clear trend from direct ($\Delta k = 0$) to indirect band gaps for nanotubes with chiral indices transitioning from $(n, 0)$ to $(0, m)$\cite{allec2016inconsistencies}. Quantum confinement and curvature systematically modify not only the band gap magnitude, but also its electronic character. Curvature-induced symmetry breaking and anisotropic strain dictate the momentum-space separation between the valence band maximum and conduction band minimum across different chiralities~\cite{fernandez2016chiral, sa2015electronic}.

The hole effective mass for 2D MLBP along the zigzag direction is remarkably heavy ($m_h^* \approx 5.0\text{--}6.0\,m_0$), whereas along the armchair direction it is exceptionally light ($m_h^* \approx 0.15\,m_0$). When rolled into a nanotube, these highly directional mass tensors remain largely preserved, resulting in a strong anisotropic transport profile. Specifically, the electron effective mass $m_e$ is low for zigzag $(n, 0)$ and ladder $(n, n)$ configurations, reaching its highest values in armchair $(0, m)$ nanotubes. Similarly, armchair PNTs exhibit the largest $m_h$ values, whereas all other chiralities display relatively small hole effective masses.

Figures~\ref{Eg-DFT}e--j present a series of band structures for nanotubes with either $n$ or $m$ equal to $10$. Starting with the armchair $(0, 10)$ nanotube, it exhibits a striking resemblance between the DFT-calculated bands and the mapped ones, featuring both a maximum and a minimum at the $\Gamma$ point alongside a significant widening of the gap toward the $X$ direction. In fact, all $(0, n)$ tubes show a strong shape correspondence between the DFT and mapped bands. Specifically, the characteristic dispersion profile of the highest valence band and lowest conduction band along the $\Gamma \to X$ path of MLBP is remarkably well preserved in the $(0, n)$ nanotubes.

In contrast, the differences between the mapped band structures and those obtained via DFT are particularly pronounced in the zigzag $(10, 0)$ PNT due to the aforementioned band gap overestimation, coupled with its true metallic nature. Furthermore, in nanotubes characterized by large supercells, such as the $(4, 10)$ and $(5, 10)$ PNTs, an exceptionally high density of bands is observed. This feature arises from zone folding into a minute one-dimensional reciprocal space, which is faithfully captured by our computational calculations.

\section{Conclusions}\label{sec3}

This work presents the first comprehensive and systematic study on the structural, stability, and electronic properties of 214 real chiral phosphorene nanotubes using a combined 2D zone-folding mapping approach and DFT calculations. 

Our findings demonstrate that the 2D mapping methodology provides an efficient first-order prediction of the overall band structure profile and subband density for some PNTs families. However, full DFT optimization is indispensable to capture the significant curvature-induced orbital rehybridization and stress relaxation effects, which substantially alter the electronic band gap compared to kinematic projections.

Structurally, small-radius PNTs experience severe radial and longitudinal strains, frequently triggering cross-sectional transitions from circular to non-circular geometries (oval, rectangular, square, and irregular forms). Thermodynamic analysis reveals a strict hierarchy of stability governed by chirality, progressing as $\text{zigzag } (n, 0) < (4m, m) < (2m, m) < (n, n) < (2n, 3n) < (n, 2n) < (n, 4n) < \text{armchair } (0, m)$, which confirms that armchair $(0, m)$ PNTs represent the lowest-energy configurations, while chiral families systematically bridge the gap between symmetric armchair and zigzag extremes.

Electronically, quantum confinement and curvature modulate both the magnitude and character of the band gap. The $(n, n)$ ladder family and general chiral configurations with angles near $45^\circ$ (\textit{e.g.}, the $(13, 10)$ PNT) exhibit the largest band gaps, whereas zigzag PNTs from $(8, 0)$ to $(15, 0)$ display metallic behavior. Furthermore, the directional effective mass tensors of monolayer phosphorene remain largely intact upon rolling, establishing highly anisotropic charge carrier transport profiles across different chiral families.

Finally, the successful generation and characterization of these 214 structures demonstrate the robust capabilities of \textit{Chiraltube} for modeling complex 2D bilayer-derived nanostructures. The ability to tailor the band gap, trigger direct-to-indirect electronic transitions, and induce metallic states through precise selection of radius and chirality establishes PNTs as versatile platforms for next-generation nanoelectronic and optoelectronic devices.

\section{Methodology}\label{sec4}
An initial set of 214 PNTs was generated from the 2D phosphorene unit cell (Figure \ref{Structures}a) using the Chiraltube construction software~\cite{de2024chiraltube} with the built in standard error of 0.05\% in the search of translational indices. Structures were defined by chiral indices $(n, m)$ within the range of $1 \leq n, m \leq 20$, excluding extremely small nanotubes with diameters below 4~\text{\AA} due to their high curvature strain. From this set, the structural parameters in real space were extracted. Subsequently, the reciprocal lattice vectors of the unfolded nanotubes were determined, establishing the corresponding cutting lines within the 2D phosphorene Brillouin zone. Finally, a 2D-to-1D zone-folding mapping was employed to evaluate the electronic structure and determine the band structure evolution of the 1D tubes.

From this pool, a representative subset of 48 nanotubes was selected for first-principles calculations based on a unit cell constraint of a maximum of 236 atoms and a radius range of $4 < r < 10$ \AA.

Total energy calculations, geometry optimizations, and electronic structure analyses were performed within the DFT framework using the Quantum ESPRESSO (QE) package~\cite{giannozzi2009quantum,giannozzi2017advanced}. Exchange-correlation effects were treated using the generalized gradient approximation (GGA) as parametrized in the Perdew-Burke-Ernzerhof (PBE) functional~\cite{perdew1996generalized}, combined with projector augmented-wave (PAW) pseudopotentials~\cite{kresse1999ultrasoft}.

Based on convergence tests for the phosphorene monolayer, a plane-wave kinetic energy cutoff of 950~eV was chosen. The Brillouin zone was sampled using a $8 \times 12 \times 1$ Monkhorst-Pack grids~\cite{monkhorst1976special} of for the 2D sheet. To eliminate artificial interactions between periodic images, a vacuum region of at least 10~\AA{} was implemented along non-periodic directions ($z$ for the monolayer; $x$ and $y$ for the nanotubes).



To ensure consistent sampling across the different chiralities and cell sizes, the Monkhorst-Pack grid was dynamically adapted to the periodic length of each nanotube. Specifically, the density of $k$-points was chosen to maintain a maximum spacing of 0.25~\AA$^{-1}$ in reciprocal space along the axial direction. For non-periodic directions, a single $k$-point was used, consistent with the large vacuum buffer employed to isolate the structures.

Both atomic coordinates and unit cell dimensions were fully relaxed using the BFGS quasi-Newton algorithm~\cite{broyden1970convergence}. Geometry optimizations were considered converged when the following criteria were simultaneously satisfied: total energy changes below $1.0 \times 10^{-5}$ Ry ($1.4 \times 10^{-4}$ eV), maximum residual forces on any atom less than $5.0 \times 10^{-5}$ Ry/Bohr ($1.3 \times 10^{-3}$ eV / \AA), and external pressure within 0.5~kbar (50~MPa).


\section{Data availability}\
The data analyzed in this study are not available as part of a public database and are available from the corresponding author on reasonable request.

\bibliography{bib}

\end{document}